\documentclass[twocolumn,english,notitlepage,superscriptaddress]{revtex4-1}
\usepackage[utf8]{inputenc}

\date{\today}

\usepackage{amsmath}
\usepackage{amssymb}
\usepackage{graphicx}
\usepackage{color}

\usepackage{babel}
\begin{document}

%\selectlanguage{english}% ?

\title{Amplification of waves from a rotating body}
%short titles are better
%rotational superradiance

\author{Marion Cromb}

\affiliation{School of Physics and Astronomy, University of Glasgow, Glasgow,
G12 8QQ, UK}

\author{Graham M. Gibson}

\affiliation{School of Physics and Astronomy, University of Glasgow, Glasgow,
G12 8QQ, UK}

\author{Ermes Toninelli}

\affiliation{School of Physics and Astronomy, University of Glasgow, Glasgow,
G12 8QQ, UK}

\author{Miles J. Padgett}
\email{miles.padgett@glasgow.ac.uk}

\affiliation{School of Physics and Astronomy, University of Glasgow, Glasgow,
G12 8QQ, UK}
\author{Ewan M. Wright}

\affiliation{College of Optical Sciences, University of Arizona, Tucson, 
Arizona 85721, USA}

\author{Daniele Faccio}
\email{daniele.faccio@glasgow.ac.uk}

\affiliation{School of Physics and Astronomy, University of Glasgow, Glasgow,
G12 8QQ, UK}
\affiliation{College of Optical Sciences, University of Arizona, Tucson, 
Arizona 85721, USA}

\begin{abstract}
 
In 1971 Zel'dovich predicted that quantum fluctuations and classical waves reflected from a rotating absorbing cylinder will gain energy and be amplified.   
This key conceptual step towards the understanding that black holes may also amplify quantum fluctuations, has not  been verified experimentally due to the challenging experimental requirements on the cylinder rotation rate that must be larger than the incoming wave frequency.
Here we experimentally demonstrate that these conditions can be satisfied with acoustic waves.
We show that low-frequency acoustic modes with orbital angular momentum are transmitted through an absorbing rotating disk and amplified by up to 30\% or more when the disk rotation rate satisfies the Zel'dovich condition.
These experiments address an outstanding problem in fundamental physics and have implications for future research into the extraction of energy from rotating systems.

\end{abstract}

\maketitle

%Nature: Abstract, Main, Methods, References
%USE STANDARD PRL FORMAT FOR NOW

{\bf{Introduction.}}
%{\bf{Introduction}}.
In 1969, Roger Penrose proposed a method to extract the rotational energy of a rotating black hole, now known as Penrose superradiance \cite{penroseGravitational1969}. Penrose suggested that an advanced civilisation might one day be able to extract energy from a rotating black hole by lowering and then releasing a mass from a structure that is co-rotating with the black hole. Yakov Zel'dovich translated this idea of rotational superradiance from a rotating black hole to that of a rotating absorber such as a metallic cylinder, showing it would amplify incident electromagnetic waves, even vacuum fluctuations, that had angular momentum \cite{zeldovichGeneration1971,zeldovichAmplification1972,zeldovichRotating1986}. These notions involving black holes and vacuum fluctuations converged in Hawking's 1974 prediction that non-rotating black holes will amplify quantum fluctuations, thus dissipating energy and eventually evaporating. Analogue laboratory experiments have been carried out that confirm these physical ideas: Penrose superradiance, {{or superradiant scattering}}, in classical hydrodynamical vortices in the form of `over-reflection' \cite{over,Silke1} and Hawking's predictions classically in flowing water \cite{silke2} and in optics \cite{Bel,Leo}, plus a quantum analogue in superfluids \cite{jeff1,jeff2,jeff3}. However, experimental verification of Zel'dovich amplification {{in the form of amplification of waves from an absorbing cylinder}} is still lacking.\\
Zel'dovich found the general condition for amplification from an absorbing, rotating body:%that if the angular wave frequency $\omega$ is smaller than the waves helical mode number $\ell$ multiplied by the absorber's angular frequency $\Omega$, i/e. %of a mode with freq and topological charge
\begin{equation}\label{e:Zeldy}
    \omega - \ell\Omega < 0
\end{equation}
where $\omega$ is the incident wave frequency, $\ell$ is the order of (what it is referred to in the current literature  as) the orbital angular momentum, OAM \cite{allenOrbital1992,andrewsAngular2013,padgettLight2004} and $\Omega$ is the rotation rate of the absorber. When this is satisfied, the absorption changes sign and the rotating medium acts as an amplifier. Outgoing waves then have an increased amplitude, therefore extracting energy from the rotational energy of the body in the same spirit of Penrose's proposal.\\
%
%generic effect - superradiance - cherenkov
%effectively the absorber needs to exceed the phase velocity of the waves
%Zel'dovich considered electromagnetic (EM) waves, amplified upon reflection from a metallic cylinder. 
Satisfying the condition in Eq.~\eqref{e:Zeldy} with electromagnetic waves is extremely challenging. For $\ell=1$ we would need rotation speeds $\Omega$ in the GHz to PHz region (microwave to optical frequencies) which is many orders of magnitude faster than the typical 100-1000 Hz rotation speeds available for motor-driven mechanically rotating objects. The highest OAM reported to date in an experiment is of order $\ell\sim10,000$ in the optical domain \cite{largeOAM}, yet still leaves little hope of closing the rotation frequency gap required to satisfy Eq.~\eqref{e:Zeldy} \cite{Ewan1,goodingReinventing2019}.\\
%
%mention the radio waves?%Perhaps super low frequency or extremely low frequency radio waves could be an option. However, these waves of 3 to 300 Hz have huge wavelengths of 1,000 to 100,000 kilometers (in comparison, the Earth itself is only about 13,000 km across). To generate such wavelengths requires extremely large antennas of at least a few km in length, and megawatts of power to operate – an experimental scale far beyond the resources of a PhD student, and really for anyone other than militaries who have already built these antennas to communicate with submarines. Although perhaps these conditions are happening out there in space somewhere!
%
However, recent work has shown that this condition and the observation of gain is theoretically achievable with acoustic waves \cite{faccioSuperradiant2019,Silke_sound,goodingDynamics2020}. The proposed interaction geometry requires  sending an acoustic wave in transmission through a rotating absorbing disk. {{This provides a strong technical advantage compared e.g. to sending the waves radially inwards towards the outer surface of a cylinder, as it allows us to use relatively low frequencies for both the waves and the disk rotation whilst keeping the dimensions sufficiently compact (the disk can be made very thin).}} An acoustic wave with OAM order $\ell$ will experience a rotational Doppler shift \cite{courtialRotational1998,bialynicki-birulaRotational1997} due to the disk rotation, such that the wave frequency is shifted by a quantity $\omega-\ell\Omega$. This implies that the acoustic wave frequency will become negative when the Zel'dovich condition Eq.~\eqref{e:Zeldy} is satisfied, which is precisely the pre-requisite physical condition outlined by Zel'dovich in his original work. This condition was recently observed by Gibson et al. \cite{gibsonReversal2018} by measuring the acoustic frequency with a rotating microphone. Although one cannot directly measure negative frequencies, the switch in sign of the acoustic wave frequency manifested itself as a switch in the sign of the wave orbital angular momentum, which was measured by tracking the phase difference between two closely spaced, co-rotating microphones.\\
In this work we experimentally demonstrate that Zel'dovich amplification is readily observable with acoustic waves with relatively low OAM ($\ell=3,4,5$) and at low acoustic frequencies of order of 60 Hz, i.e. readily accessible rotation rates for the absorbing disk such that spurious signals (for example, due to noise) are also minimised. Our acoustic measurements are resolved as a spectrogram and analysed as a function of disk rotation frequency, showing an intensity gain of $\sim$30\% of acoustic energy over a range of orbital angular momenta. These measurements represent a significant step forward in our understanding of Zel'dovich amplification, a fundamental wave-matter interaction that lies at the heart of a series of physical processes in condensed matter systems, superfluids and black holes.\\
\begin{figure}[tb]
\includegraphics[width=1\columnwidth]{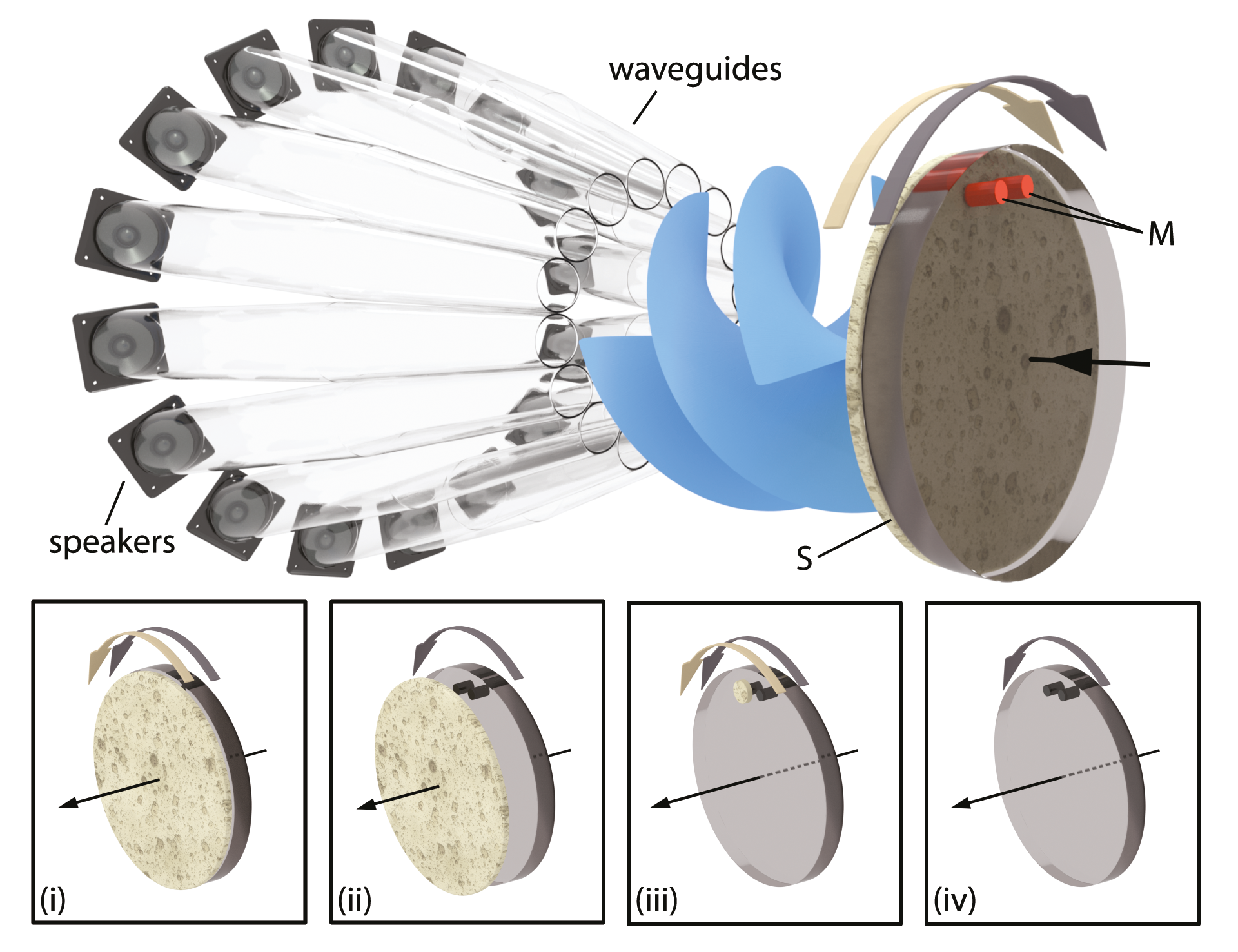}%png should be a pdf or svg!
\caption{{\bf{Schematic outline of experiment.}} 16 loudspeakers (Visaton; SC 8N) are arranged in a ring (diameter $\approx$0.47m) to create an OAM acoustic field, channelled by acoustic waveguides to a smaller area (diameter $\approx$0.19 m) and incident on a rotating disk of sound-absorbing foam (S). The absorbing disk also carries two closely spaced (2 cm distance) microphones (M). The microphones transmit their data via Bluetooth (Avantree; Saturn Pro), for live data acquisition whilst in rotation. The set-up is adapted from that used by Gibson et. al. \cite{gibsonReversal2018}. Insets indicate the various configurations used in the experiments for the rotating disk and absorbing foam: (i) supporting disk with microphones and absorber are co-rotating; (ii) absorber is detached and remains static, whilst microphones rotate; (iii) an absorber is placed in front of only one of the two microphones; (iv) absorber is completely removed, microphones rotate.}
\label{f:setup}
\end{figure}
{\bf{Model.}}
An acoustic conical wave carrying OAM $\ell$ is normally incident onto an absorbing material rotating at frequency $\Omega$, and which is surrounded on both sides by non-rotating air. The acoustic wave equation for density variations $\tilde\rho$ in a frame rotating with the medium is \cite{boyd}:
\begin{equation}\label{start}
{\partial^2 \tilde\rho\over \partial t^2} - \Gamma'\nabla^2 {\partial\tilde\rho\over \partial t} - v^2 \nabla^2   \tilde\rho = 0,
\end{equation}
where $v$ is the sound velocity and $\Gamma'$ the damping parameter:  A similar wave equation applies in the surrounding air with sound velocity $v_0$ and $\Gamma'=0$.
\begin{figure}[!t]
\includegraphics[width=8cm]{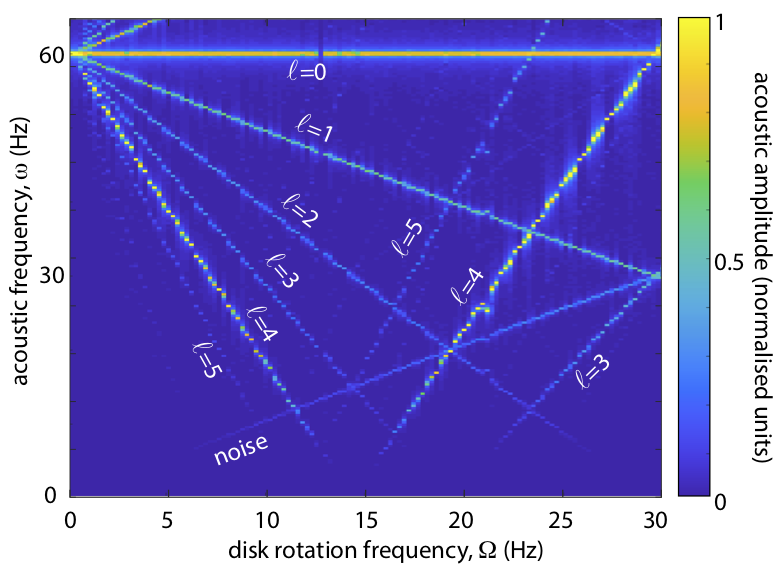}%png should be a pdf or svg!
\caption{{\bf{Spectrally resolved acoustic measurements.}} An example of a measured spectrogram showing the measured acoustic frequency ($\omega$) spectrum in the rotating frame for increasing rotation frequencies ($\Omega$). The OAM beam is generated in the lab frame at 60 Hz at a constant volume, with the speaker output phases optimised for the $\ell$=4 mode. For each value of $\Omega$, the data shows an independent spectrum, obtained from the Fourier transform of the measured signal from one of the two microphones on the rotating disk. The data clearly shows the input 60 Hz signal split into multiple components, corresponding to the various OAM modes (indicated in the graph) as a result of a rotational Doppler shift, $\omega-\ell\Omega$, that leads to linearly varying frequency as a function of $\Omega$, for each $\ell$-mode. The microphone response decreases for decreasing measured $\omega$ and below 4 Hz is zero (i.e. below the noise level). The supplementary video shows an animation with the overlaid acoustic signal that is recorded with increasing $\Omega$.}
\label{f:spectrogram}
\end{figure}
Under the condition that the medium length $L$ is much less than the acoustic wavelength, the transmission of the  beam incident from air onto the rotating medium may be solved by treating the effects of the medium absorption term in \eqref{start} within the first Born approximation. The details of this model have been worked out previously (supplementary material in \cite{faccioSuperradiant2019}) and lead to the following expression for the acoustic beam transmittance:
\begin{equation}\label{theory}
T=\left[1-\cfrac{L\omega^2}{k_z v^4}\Gamma' (\omega-\ell\Omega)\right]C(\omega),
\end{equation}
where $k_z=(\omega/v)\cos\theta$ is the longitudinal component of the sound wavevector and can be controlled through the conical beam focusing angle, $\theta$. We underline that it is the term $(\omega-\ell\Omega)$ in the transmittance that can change the sign of the absorption and lead to gain in correspondence with the Zel'dovich condition in Eq.~\eqref{e:Zeldy}. \\
{{Equation~\eqref{theory} also includes the frequency response of the microphones, $C(\omega)$}}. %These are placed on the rear of the rotating disk (behind the absorber) and therefore measure a Doppler shifted frequency. 
Standard microphones exhibit a roll-off in sensitivity starting below $\sim100$ Hz. We model this response with a function $C(\omega)=1-\exp[-(\omega-\ell\Omega)^2/\sigma^2]$, where $\sigma$ determines the rate at which the sensitivity drops as a function of frequency. However the precise form of this function is not critical to our main conclusions, as the experiments described below compare between two microphones with the same frequency response.\\%, and even with a single microphone we can cancel out this effect by comparing points on the OAM traces which have the same measured acoustic frequency before and after the Zel'dovich condition (Eq.~\eqref{e:Zeldy}) is satisfied.\\
{\bf{Experiments.}}
We generate an acoustic wave with orbital angular momentum using a ring of speakers and tubes that guide the sound directly onto the rotating disk, as shown in Fig.~\ref{f:setup}. The ring of 16 loudspeakers are all driven at the same frequency ($\omega = 60$ Hz), each with a specific phase delay in order to approximate a helical phase front, generating a beam carrying OAM \cite{gibsonReversal2018}. Depending on the phase delay between adjacent speakers, different OAM states can be produced. For example, a phase delay of $\pi/2$ radians between adjacent speakers creates an OAM beam of topological charge $\ell=4$.\\
\begin{figure}[!t]
\includegraphics[width=7cm]{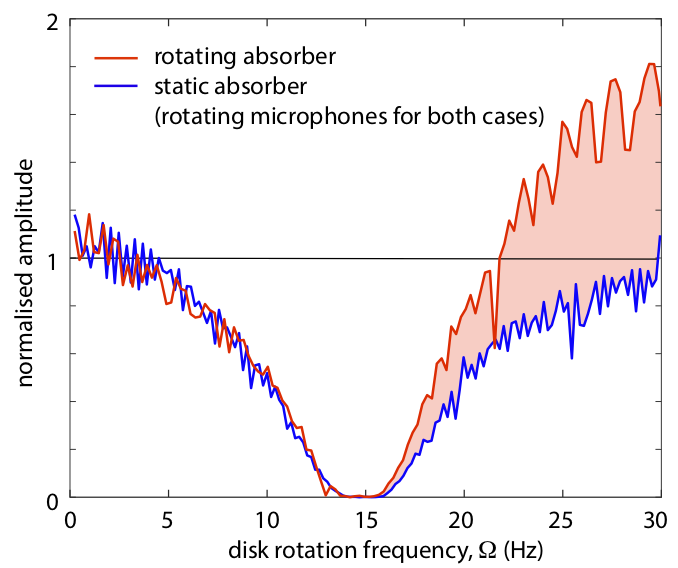}
\caption{{\bf{The effect of rotation.}} A measurement of the acoustic amplitude for $\ell=4$ for the case of a rotating absorber (red curve) and for the case in which the absorber is detached from the rotating disk holding the microphones, and hence is static (blue curve). The rotating absorber case shows a clear increase of the transmitted acoustic amplitude above the Zel'dovich condition ($\omega-\ell\Omega<0$ is satisfied for this case when $\Omega>15$ Hz). }
\label{f:staticabsorber}
\end{figure}
A motor (RS Components; 536-6046) is used to rotate the disk fitted with two closely spaced microphones. Sound absorbing material can be placed in front of both, one or neither of the microphones (as illustrated in Fig.~\ref{f:setup} (i), (iii) and (iv) respectively). Test measurements are taken with the two microphones under experimental conditions in order to ensure that they exhibit the same acoustic response, with and without the absorbing material placed in front of them (see Methods). The data from the microphones is communicated via Bluetooth to a computer.\\% for Fourier analysis. \\
Fig.~\ref{f:spectrogram} shows an example of a measured spectrogram. The acoustic frequency is set to 60 Hz on all of the speakers and phase delays are set to generate waves with $\ell=4$ - other $\ell$ modes are expected to also be generated as a result of the imperfections in speaker uniformity and the limited number of speakers used \cite{gibsonReversal2018}. The Zel'dovich condition and inversion from absorption to gain is therefore expected for a disk rotation of 15 Hz. The disk is therefore rotated in the 0 to 30~Hz range, which also corresponds to the linear response range of our motor (i.e. linear increase of rotation speed with driving voltage). The spectrogram exhibits a series of features: as the disk rotation rate increases, the input 60 Hz frequency splits into a series of signals, depending on the OAM value $\ell$ with a clear signal measured for $\ell=0-5$ (as labelled in the figure). We can also clearly see an additional signal that is due to the noise generated by the rotation and therefore appears at the same frequency as the rotation rate. \\
\begin{figure}[!t]
\includegraphics[width=7cm]{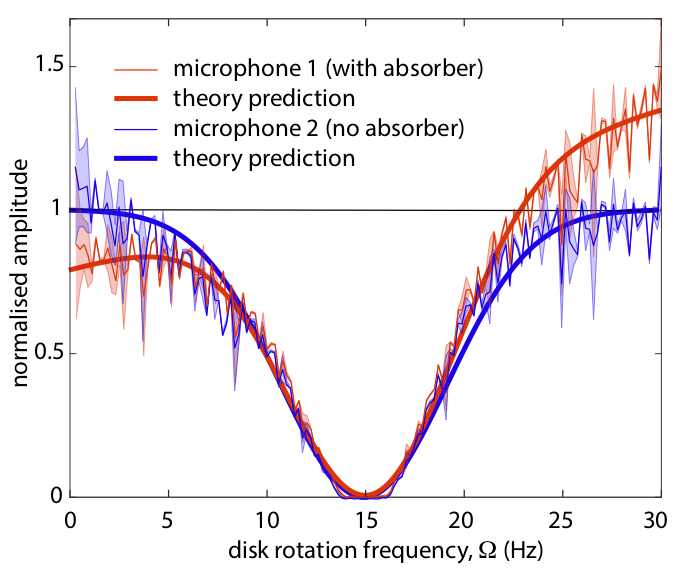}
\caption{{\bf{Evidence of absolute gain.}} The measured acoustic amplitude with $\ell=4$ and with the absorber placed on one of the microphones (red curve) but not on the other (blue curve) shows clear differences in the signals.  For rotation rates $\Omega<15$ Hz (i.e. such that $\omega-\ell\Omega>0$)  absorption is observed in particular at the lowest frequencies (blue shaded area). Conversely, at the highest frequencies (where $\omega-\ell\Omega<0$), a clear gain in the transmitted signal is observed. The $\sim1.3$x higher signal at $\Omega\sim30$ Hz compared to $\Omega\sim0$ Hz highlights the presence of absolute gain of the acoustic signal.  Theoretical predictions from Eq.~\eqref{theory} are shown with the damping parameter $\Gamma'=0$  m$^2$/s (no absorber, thick blue curve) and $\Gamma'=8\cdot 10^4$ m$^2$/s (%absorber of thickness L=1 mm, 
thick red curve) {{The shaded areas show the standard deviation of the measured signals across 7 sets of data (2 seconds of acquisition each)}}.
}
\label{f:foamnocomp}
\end{figure}
All of the observed OAM modes shift in frequency due to the rotational Doppler shift ($\Delta\omega = -\ell\Omega$) and after the labelled OAM modes have gone through zero frequency they satisfy their Zel'dovich condition. Beyond zero, the rotational Doppler shift formula predicts negative frequencies, which results in an inversion of the sign of $\ell$ (i.e. positively sloped traces in the spectrogram) when measured in the rotating frame  \cite{gibsonReversal2018}. In order to verify the presence of gain in this Zel'dovich regime, we proceed to extract the amplitude for each $\ell$ value from the spectrograms. \\
\begin{figure}[!t]
\includegraphics[width=7cm]{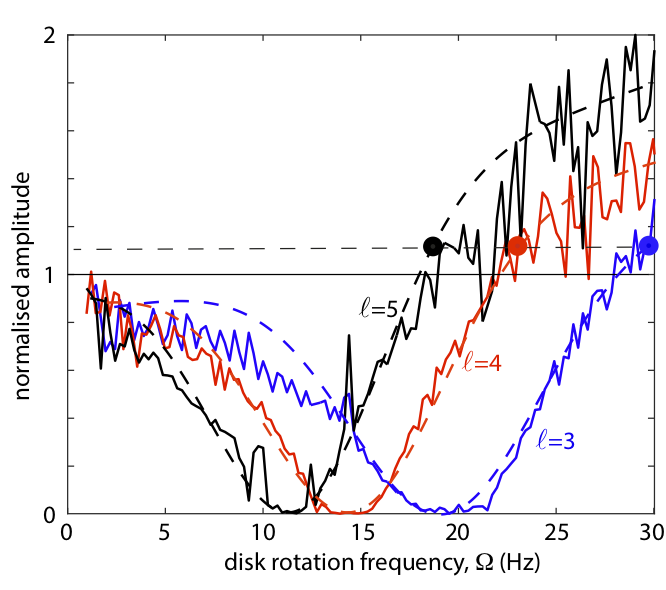}
\caption{{\bf{Comparison for different OAM beams.}} The spectrograms show signals for a range of $\ell$ that can also be analysed and compared. For all $\ell$ that pass through the Zel'dovich condition we see evidence of transmittance greater than 1 as a result of rotation. Comparing the transmission values for the same rotational Doppler shifted frequency (i.e. -30 Hz, corresponding to $\Omega=30$, 22.5 and 18 Hz (indicated as solid circles) for $\ell=3$, 4 and 5, respectively), the gain in transmittance appears, within the experimental error, to be constant, $\sim1.1$ (horizontal dashed line) for all $\ell$ and increases linearly with $\ell$ for a fixed $\Omega$ (compare e.g. at $\Omega=30$ Hz). Both observations confirm the predictions of Eq.~\eqref{theory}. Theoretical fits from Eq.~\eqref{theory} are also shown as dashed lines, with no varying parameters (other than $\ell$).
}
\label{f:multipleOAMamplified}
\end{figure}
{\bf{Results.}}
%\textcolor{red}{corrected up to here}
%
In Fig.~\ref{f:staticabsorber} we show the effect of rotation on transmitted acoustic signal for the $\ell$=4 mode as the disk rotation rate is increased from 0 to 30 Hz. The two curves indicate two different cases:  the absorbing disk is co-rotating with the microphones (red curve) and the absorbing disk is slightly detached from the motor mount so that the microphones rotate whilst the disk remains static (blue curve). As the rotation speed is increased, the modes are Doppler shifted and the measured signal from both microphones decreases due to the lower microphone response at lower acoustic frequencies. As the mode is Doppler shifted through zero frequency (at $\Omega=15$ Hz), the measured acoustic frequency increases again and the transmitted signal increases. In the non-rotating case, no increase is observed in the transmitted signal for the same rotational Doppler shift (i.e. for symmetric points around $\Omega=15$ Hz). Conversely, when the absorber is in rotation with no other changes to the experiment, we observe a clear increase in the transmitted signal at high rotation rates that satisfy the Zel'dovich condition (shaded area).\\
In Fig.~\ref{f:foamnocomp} we show evidence of absolute gain in the acoustic signal, i.e. evidence that the transmitted energy is larger than the incident energy.
One microphone (microphone 1, red curve) in the rotating frame has absorbing foam in front of it, the other microphone next to it (microphone 2, blue curve) does not. We observe that at low rotation speeds (2-5 Hz), the transmitted signal is lower compared to microphone 2, as it has been absorbed by the foam. Conversely, rotating faster than $\Omega=15$ Hz and thus satisfying the Zel'dovich condition leads to clear increase in the transmission signal compared to the non-absorbing case.
The amplification is such that the signal transmitted through the absorber above $\Omega=25$ Hz is greater by about 30$\%$ than the signal at the slowest rotation speeds that did not pass through the absorber. This indicates absolute gain: we measure more sound with the rotating absorber than without it. \\
The thick solid curves in Fig.~\ref{f:foamnocomp} show the theoretical predictions from Eq.~\eqref{theory}. We first proceed to fit Eq.~\eqref{theory} to the data from microphone 2 that has no absorber present ($\Gamma'=0$) thus obtaining the shape of $C(\omega)$, the frequency response of the microphones. This determines the frequency sensitivity parameter, $\sigma=22$ Hz. We then refer to the data from microphone 1 and use the low rotation frequency (2-5 Hz) data to determine the value of the dissipation parameter, $\Gamma'=8\cdot 10^4$ m$^2$/s. % This value agrees to within a factor $\sim2$x with that measured independently with a commercial sound-meter. 
We notice that the same theoretical curve provides a quantitatively accurate prediction of  the full behaviour for all $\Omega$, including the 30\% gain measured at high rotation frequencies, thus further corroborating the interpretation of the gain originating from the Zel'dovich effect.\\
Further analysis shown in Fig.~\ref{f:multipleOAMamplified} of multiple OAM modes transmitted through the rotating absorber reveals amplification for all the OAM modes analysed that satisfy the Zel'dovich condition, not just the strongest $\ell=4$ mode. In more detail, if we consider in Fig.~\ref{f:multipleOAMamplified} a fixed Doppler shifted frequency, e.g. $\omega-\ell\Omega=-30$ Hz for all $\ell$ (corresponding to $\Omega=30$, 22.5 and 18 Hz for $\ell=3$, 4 and 5, respectively), we note that all curves within the experimental error show the same gain of $\sim10\%$. If instead we consider a fixed disk rotation frequency, e.g. $\Omega=30$ Hz we see that the gain, i.e. transmitted acoustic energy, increases linearly with $\ell$. Both of these observations are in agreement with the theoretical prediction Eq.~\eqref{theory}. \\%We note that we perform these comparisons for the highest value of $\Omega$ in order to minimise the role of the microphone response which flattens out above 40-50 Hz acoustic frequencies.\\
{\bf{Conclusions.}}
Amplification of waves from a rotating absorber as predicted by Zel'dovich is a foundational prediction in fundamental physics that lies somewhere between the proposition by Penrose that energy can be extracted from rotating black holes and Hawking's prediction that static black holes will evaporate as a result of the interaction with quantum vacuum. Zel'dovich's original model indeed referred to amplification of vacuum modes from a rotating metallic cylinder but was also extended to include the amplification of classical waves. Whilst very hard to verify with optical or electromagnetic waves, acoustic waves allow direct measurements of significant amplification of waves due to a rotating absorber. A key step in achieving this result is the use of a geometry where the waves are transmitted through a thin absorbing cylinder \cite{faccioSuperradiant2019,Silke_sound,goodingDynamics2020} rather than in reflection from an extended cylinder. This relaxes the experimental constraints and limitations that arise in the original proposal due to the requirement that the cylinder length be larger than the wavelength, in order to ensure interaction and reflection of the incident waves. For example, this would have required a cylinder with a length of several meters for the conditions used here, which would have been very challenging to rotate at 30 Hz.\\
Similar concepts could in principle be extended to electromagnetic waves \cite{goodingReinventing2019} %(although the precise operating conditions remain to be fully identified)
thus possibly extending our results also to amplification of electromagnetic modes from the quantum vacuum.\\

{\bf{Acknowledgements.}}
The authors acknowledge financial support from EPSRC (UK Grant No. EP/P006078/2) and the European Union's Horizon 2020 research and innovation programme, grant agreement No. 820392.\\
%All data used in the measurements presented in this work is available from DOI-to-be-provided.

{\bf{Author contributions.}}
MC, GMG performed the measurements and data analysis. GMG, ET, MC prepared the experiment. EMW, DF and MJP conceived the experiment and theory. All authors contributed to the manuscript.
\\

\begin{figure}[!t]
\includegraphics[width=5cm]{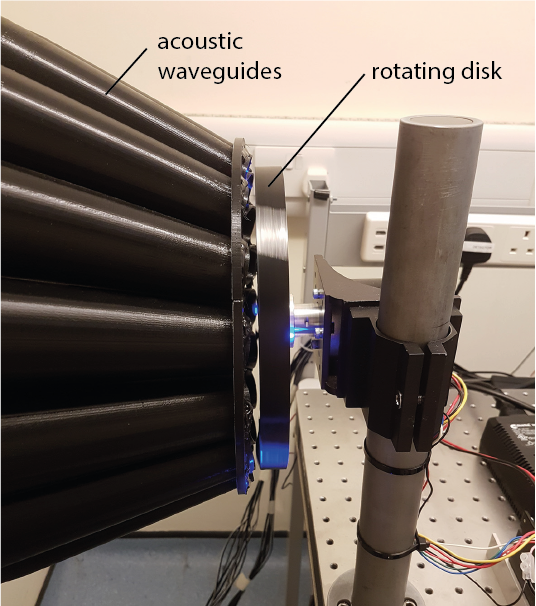}
\caption{{\bf{Photograph of the setup}} showing the detail of the interaction region where the acoustic waveguides conduct the sound directly on to the absorber, supported by a plastic disk.
}
\label{f:photo}
\end{figure}

{\bf{Methods.}}
%\subsection{Analysis}
The rotation speed of the absorber was increased in steps of (approximately) 0.2 Hz. Various forms of sound absorbing foam were tested with varying yet similar porosity {{(e.g. cellular ethylene propylene diene monomer, EPDM,  soundproofing rubber, RS Components, 5\% absorption at 60 Hz)}}. All cases showed similar results, in line with our expectation that details in the medium 4-5 orders of magnitude smaller than the sound wavelength will not significantly influence the dynamics.\\
{{Figure~\ref{f:photo} shows a photograph of the sound-absorber interaction region. The acoustic waveguides can be seen on the left, conducting the sound towards and directly on to the rotating absorber. The absorbing foam is held in place with a support structure, which is made of a plastic disk with no air gaps or possibility for sound to reach the microphones (5 mm diameter, embedded in the supporting plastic disk) without passing through the foam. This setup ensures that all sound reaches the microphones only through the foam.}}\\
%
%An example is shown in Fig.~\ref{foam}.\\
%
%\begin{figure}[!t]
%\includegraphics[width=3cm]{foam.png}
%\caption{{\bf{Example of sound-absorbing foam:}} An example of the foam used as absorbing material in  the experiments. The full disk is shown here (diameter 20 cm, thickness 1 cm) 
%\label{foam}
%\end{figure}
%
%
\begin{figure}[!t]
\includegraphics[width=4cm]{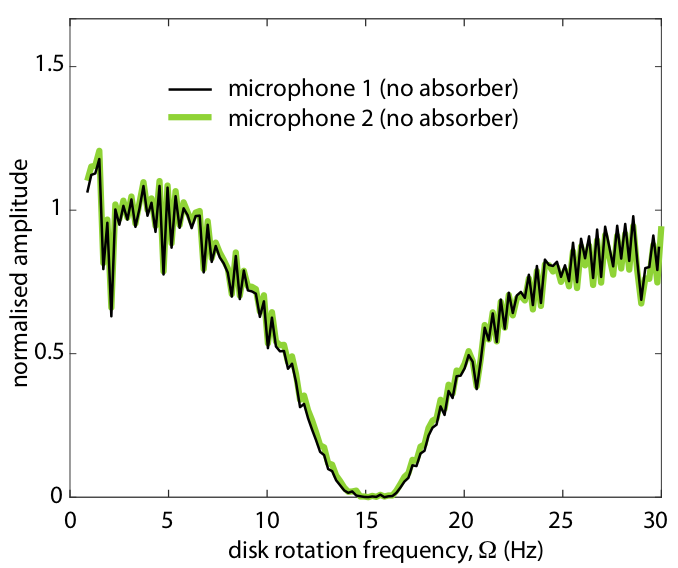}
\includegraphics[width=4cm]{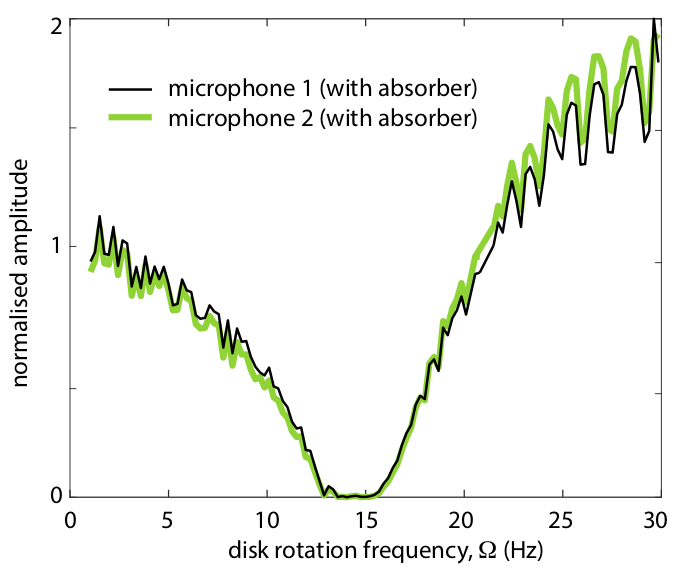}
\caption{{\bf{Microphone calibration:}} measurements of response when both microphones have no absorber or both have absorbers placed in front of them, showing that the microphones ar both calibrated and measure, as desired, the same signal. 
}
\label{f:compare}
\end{figure}
For each rotation speed, sound was recorded for short time intervals, e.g. 2 to 3 seconds. The microphone signal was then Fourier transformed (and )averaged over 2 to 3 separate measurements) so as to decompose the signal into its frequency spectrum. The frequency spectrum for each rotation speed was then used to create a single matrix of the full spectrogram (e.g. Fig.~\ref{f:spectrogram}). In MATLAB the `tfridge' range of functions was used to extract the signal amplitude (in arbitrary units) along each OAM mode in this spectrogram. The highest neighbouring frequency bin for each rotation speed was added to the signal in order to reduce noise from the discretisation of the Fourier-transformed data. \\
We also verified that the two microphones in our setup are calibrated so as to provide the same response for the same incident signal, for all rotation speeds. Fig~\ref{f:compare} shows two graphs with measurements of the two microphone responses (black and green curves) when both are uncovered (absorber removed) or both have an absorber placed in front of them. Both graphs show a nearly identical response for the two microphones under the operating conditions of our experiments.

\end{document}